\documentclass{ws-p8-50x6-00}

\usepackage{amsfonts,epsfig,mathbbol}

\def\D{{\rm D}}

\begin{document}

\title{Pointlike Hopf Defects in Abelian Projections}

\author{Falk Bruckmann}

\address{Theoretisch-Physikalisches Institut,
Friedrich-Schiller-Universit\"at Jena\\
Max-Wien-Platz 1, D--07743 Jena\\
email: {\tt brk@tpi.uni-jena.de}}

\maketitle

\abstracts{
We present a new kind of defect in Abelian Projections,
stemming from pointlike zeros of second order.
The corresponding topological quantity is the
Hopf invariant $\pi_3(S^2)$ (rather than the winding number
$\pi_2(S^2)$ for magnetic monopoles). We give a visualisation
of this quantity and discuss the simplest non-trivial example, the
Hopf map.
Such defects occur in the Laplacian Abelian gauge in a non-trivial
instanton sector. For general Abelian projections we
show how an ensemble of Hopf defects accounts for the instanton number.}

\section{Introduction}

It has long been speculated that confinement in pure Yang-Mills
theories may be realised via dual superconductivity\cite{thmapana}. 
To arrive at this picture, 't Hooft has
suggested to use Abelian projections\cite{th}. 
In this technique one fixes the gauge group up to its maximal Abelian subgroup.
The generic defects of this gauge fixing are magnetic monopoles.
They are supposed to play the role of dual Cooper pairs and force the 
chromoelectric field into strings.

Abelian projections are best described by an `auxiliary Higgs field'
$\phi$ in the adjoint representation. 
We will focus on the gauge group $SU(2)$ in the following.
An Abelian gauge (AG) assigns to every field $A$ a field $\phi$
in such a way that
{\em the gauge transformation $\Omega$ which diagonalises} $\phi$
{\em is the one which brings $A$ into the AG}.
The unfixed $U(1)$ consists of
rotations around the 3-direction in isospace. Defects of such a
gauge fixing arise when $\phi$ vanishes at some $\bar{x}$:
$\Omega$ is not well-defined there. For a topological description
define $n\equiv\phi/|\phi|$ around $\bar{x}$, which then is a regular
mapping onto $S^2\cong SU(2)/U(1)$.

\section{Zeros of the Higgs field}

There are different types of zeros of $\phi$. Generically, this field
vanishes on lines/loops, since there are three equations to be
solved on a four dimensional manifold. As an example
consider  $\vec{\phi}=\vec{x}$ with a first order zero.
This `hegdehog' field produces a static defect at the origin
of $\mathbb{R}^3$. The associated
$n$ is a mapping $S^2_{|\vec{x}|\,\rm fixed}\rightarrow S^2$, which is
characterised by an integer winding number, $\pi_2(S^2)=\mathbb{Z}$.
It counts how many times one sphere is covered by the other
and equals the magnetic charge of a monopole arising
in the Abelian projected gauge field. 
When one performs the diagonalisation of $\phi$,
a Dirac string piercing any sphere around
the defect is unavoidable.

Alternatively $\phi$ may vanish on isolated points. 
Consider $|\phi|=r^2$ which has a zero of second order
at the origin. 
Now $n: S^3_{r\,\rm fixed}\rightarrow S^2$ gives rise to another topological
quantity, the Hopf invariant, $\pi_3(S^2)=\mathbb{Z}$.

\section{The Hopf invariant}

The definition of the Hopf invariant via the homotopy group
is rather abstract.
A more intuitive form can be used under some regularity assumptions
\cite{du}. 
Then the preimage of a point on $S^2$ is a loop on $S^3$.
The Hopf invariant counts {\em how many times two such loops are linked}. 
This linking number can be understood via a combination of
Biot-Savart's and  Ampere's law \cite{fr}.

The topology of $n$ and its 
diagonalising gauge transformation $\Omega$ are related. 
The latter is a mapping $S^3\rightarrow S^3\cong SU(2)$
with the usual winding number 
equal to the Hopf invariant.
By encoding the topology in $\Omega$ it is possible
to diagonalise $n$ smoothly,
there are no further Dirac strings.

\section{The Hopf map}

\begin{figure}[!t]
\centerline{\epsfig{file=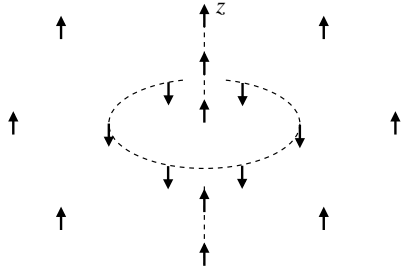, scale=0.5}}
\caption{The Hopf map visualised as a system of spins with normalised
length ($S^2$) in compactified space ($\dot{\mathbb{R}}^3 \simeq  S^3$).}
\label{soliton}
\end{figure}

A nice example of a non-trivial mapping $n$ is the Hopf
map\cite{na}. Take
\begin{equation}
\phi_a=\left(\begin{array}{c}
2(x_1x_3+x_2x_4)\\
2(x_2x_3-x_1x_4)\\
x_1^2+x_2^2-x_3^2-x_4^2
\end{array}\right),\quad |\phi|=r^2,\quad n=\phi/r^2\equiv n_{\rm H}.
\end{equation}
It has a second order zero at the space-time origin and the
corresponding $n$ has Hopf invariant one. In order to visualise the
latter, one views $S^3$ as compactified $\mathbb{R}^3$
(cf.~Fig.~\ref{soliton}).
The preimages of the north and south pole are the $z$-axis and a circle in the 
$xy$-plane, respectively, being linked once.
The energy density 
is proportional to the gradient of the spins,
thus here it is localised near that circle.
The configuration plays a role in a Skyrme-like model for low energy
QCD, where it is called {\em torus-shaped unknot soliton}\cite{fani}.

This `standard' mapping is diagonalised by another `standard' mapping
$\tilde\Omega=(x_4\Eins+ix_a\sigma_a)/r$, the identity $S^3\rightarrow
S^3$.

\section{Significance for Abelian projections}

\begin{figure}[!t]
\centerline{\epsfig{file=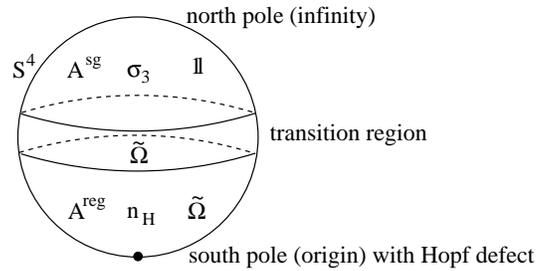, scale=0.5}}
\caption{The one-instanton one the sphere, its Higgs field and
diagonalising gauge transformation in the LAG.}
\label{bundle}
\end{figure}

We recently found Hopf defects in the
Laplacian Abelian gauge (LAG)\cite{br1}.
In this gauge the Higgs field $\phi$ is defined as the ground state 
of the covariant Laplacian $-\D^2[A]$ in the background of $A$.
In order to avoid a pure scattering spectrum, one better works in a
finite volume space-time say the sphere or the torus.
 On $S^4$ the fibre bundle construction of the 't~Hooft
instanton consists of two patches around the north and south pole with
$A$ in singular and regular gauge, respectively (cf.~Fig.~\ref{bundle}).
Inbetween 
they are related by the gauge transformation $\tilde\Omega$.
Demanding $\phi$ to have the same
transition function $\tilde\Omega$,
$n$ comes out to be $\sigma_3$ and $n_{\rm H}$
over these patches, which gives a Hopf defect at the instanton core
(the origin). In order to diagonalise $n$ one has to apply
$\tilde\Omega$, which also transforms $A^{\rm reg}$ into $A^{\rm sg}$.
The result is the $A$ field in
singular gauge everywhere\footnote{like in the maximal Abelian gauge}.

The Hopf defect may also be seen as a (twisted) {\em monopole loop
with vanishing radius}.
Thus a generic perturbation of the instanton induces a
monopole loop again\cite{br2}.
In the same manner one can understand the occurence of two monopole loops for 
the instanton-anti-instanton in the LAG\cite{reto}.

For general Abelian gauges the discussion of defects is based on the
Hopf invariant\cite{ja}: Any localised defect (points, loops or else)
can be enclosed by a sphere.
There $n$ and its Hopf invariant can be computed.
Like in residue calculus, a sphere containing no defect
cannot carry a Hopf invariant. Thus the signed sum of all Hopf
invariants gives the Hopf invariant at the boundary/in the transition
region, which is exactly the instanton number,
\begin{eqnarray}
\sum_i \mbox{Hopf}|_{S^3_i}(n)=\mbox{instanton number}(A).
\end{eqnarray}
That is, in a non-trivial background there {\em must} be
defects. However, some of the defects may cancel in the sum.
For the instanton in the LAG only the minimal number of defects arises.\\

\noindent\textbf{Acknowledgements}\\

\noindent I thank the organisers for arranging a very stimulating school.


\begin{thebibliography}{99}
\bibitem{thmapana}
Y.~Nambu, \Journal{Phys.~Rev.}{D10}{4262}{1974},
G.~Parisi, \Journal{Phys.~Rev.}{D11}{970}{1975},
G.~'t~Hooft, in: {\em High Energy Physics},
Editrice Compositori (Bologna, 1976),
S.~Mandelstam, \Journal{Phys.~Rep.}{23}{245}{1976}.
\bibitem{th} G.~'t~Hooft, \Journal{Nucl.~Phys.~B}{190}{455}{1981}.
\bibitem{du} B.~A.~Dubrovin, A.~T.~Fomenko, S.~P.~Novikov,
             {\em Modern Geometry -- Methods and Applications}
             (Springer, 1985).
\bibitem{fr} T.~Frankel, {\em The Geometry of Physics}
             (Cambridge University Press, 1997).
\bibitem{na} H.~Hopf, \Journal{Math.~Ann.}{104}{637}{1931},
             M.~Nakahara, {\em Geometry, Topology and Physics}
             (Adam Hilger, 1990).
\bibitem{fani} L.~Faddeev, A.~Niemi, hep-th/9705176, and references
               therein.
\bibitem{br1} F.~Bruckmann, T.~Heinzl, T.~Vekua, A.~Wipf,
              \Journal{Nucl.~Phys.~B}
              {593}{545}{2001}, hep-th/0007119.
\bibitem{br2} F.~Bruckmann, hep-th/0011249.
\bibitem{reto} H.~Reinhardt, T.~Tok, hep-th/0009205.
\bibitem{ja} O.~Jahn, \Journal{J.~Phys.~A}{33}{2997-3019}{2000},
             hep-th/9909004.


\end{thebibliography}
\end{document}